\documentclass[aps,prc,preprint,superscriptaddress,amsmath,showpacs,amssymb]{revtex4-1}
\usepackage{bm,physics}

\usepackage[dvips,final]{graphicx}

\usepackage{ulem, color}
\usepackage{hyperref}

\bmdefine{\ba}{a}
\bmdefine{\bb}{b}
\bmdefine{\bx}{x}
\bmdefine{\by}{y}
\bmdefine{\bz}{z}
\bmdefine{\bn}{n}
\bmdefine{\bp}{p}

\newcommand{\BM}{\begin{pmatrix}}
\newcommand{\EM}{\end{pmatrix}}

\renewcommand{\d}{\dagger}

\newcommand{\Lc}{\mathcal{L}}
\newcommand{\Mc}{\mathcal{M}}

\newcommand{\hphi}{\hat\varphi}
\newcommand{\hpsi}{\hat\psi}

\newcommand{\intx}{\int\! d^3x\;}
\newcommand{\intxd}{\int\! d^3x'\;}
\newcommand{\intxxd}{\int\! d^3x\,d^3x'\;}

\newcommand{\ex}{\mathrm{ex}}

\begin{document}

\title {Supersolidity of  the
  $\alpha$ cluster  structure in the nucleus   $^{12}$C
}
%
\author{S.~Ohkubo}
\email{shigeo.ohkubo@rcnp.osaka-u.ac.jp}
\affiliation{Research Center for Nuclear Physics,
Osaka University, Ibaraki,
Osaka 567-0047, Japan}

\author{J.~Takahashi}
\affiliation{Department of Electronic and Physical Systems,
Waseda University, Tokyo 169-8555, Japan}

\author{Y.~Yamanaka}
\affiliation{Department of Electronic and Physical Systems, Waseda University, Tokyo 169-8555, Japan}

\begin{abstract}
 For more than  half a century, the   structure of $^{12}$C, such as the ground band, has been  understood  to be well described by the  three $\alpha$ cluster model    based on a   geometrical crystalline picture.
On the contrary,  recently it has been claimed that the   ground state of $^{12}$C is  also well  described  by a  nonlocalized cluster model without any geometrical configurations originally proposed to explain  the  dilute gas-like Hoyle state,  which is now considered to be a Bose-Einstein condensate  of  $\alpha$ clusters.
  The  challenging unsolved   problem is how  we can  reconcile the two exclusive $\alpha$ cluster pictures of $^{12}$C,  crystalline  vs  nonlocalized  structure.
We show that   the crystalline cluster picture and the  nonlocalized cluster picture can be  reconciled  by noticing that they are a manifestation of supersolidity with   properties of both crystallinity  and   superfluidity.
 This is achieved through a  superfluid $\alpha$ cluster model based on   effective  field theory, which treats the Nambu-Goldstone  zero-mode rigorously.
  For several decades,  scientists have been searching for a supersolid in nature.
   Nuclear $\alpha$ cluster structure is  considered to be the first confirmed example of a stable supersolid.
  \end{abstract}
\maketitle
\par

\par
A  supersolid \cite{Andreev1969,Chester1970,Leggett1970,Matsuda1970,Boninsegni2012,%
Yamamoto2013,Leonard2017,Li2017,Tanzi2019,Bottcher2019,Chomaz2019} is  a solid that exhibits the  property of  superfluidity.  Supersolids have been searched for in He II
and recently  in the Bose-Einstein condensate (BEC) of  atomic gas.
{Superfluidity is caused by spontaneous symmetry breaking (SSB) of the global phase
and has been  known even for  small systems with
ten or fewer interacting and trapped particles such as  parahydrogen \cite{Kuyanov2008,Khairallah2007,Li2010} and He clusters \cite{Paesani2005,McKellar2006}.  In nuclei, it has been  shown that the interacting  gas-like  three  $\alpha$ clusters in the  $0_2^+$ (7.65 MeV) Hoyle state of $^{12}$C is a BEC   \cite{Tohsaki2001,Funaki2003,Matsumura2004,Itoh2004,Itoh2011,Freer2011,Nakamura2016,Nakamura2018,Katsuragi2018} and  a superfluid cluster model
 reproduced the experimental data well.
}
In the present paper, we show that  crystalline $\alpha$ cluster structure,
 which has been described  successfully by many  cluster models, has the  simultaneous properties   of crystallinity  and superfluidity. That is, the    $\alpha$ cluster structure is a stable supersolid.

\par
  The geometrical  $\alpha$ cluster model based on the crystalline picture \cite{Brink1966,Ikeda1968,Brink1968,Brink1970}, which was  originally proposed following the  intuitive  geometrical classical picture \cite{Wefelmeier1937,Wheeler1937,Wheeler1937B}, has witnessed much   success in recent  decades
   \cite{Suppl1972,Wildermuth1977,Suppl1980,Suppl1998,Tamagaki1969,Ohkubo2016}.
 This model  has explained the structure of light  nuclei\cite{Suppl1972,Wildermuth1977,Suppl1980} (typically $\alpha$+$\alpha$ cluster structure in $^8$Be and $\alpha$+$^{16}$O cluster structure in $^{20}$Ne) and  in medium-weight and  heavy nuclei \cite{Suppl1998} (typically $\alpha$+$^{40}$Ca structure in $^{44}$Ti and $\alpha$+$^{208}$Pb structure in $^{212}$Po).
  The  emergence of cluster  structure is found to be   a consequence of the Pauli principle \cite{Tamagaki1969,Ohkubo2016}.

\par
  The $\alpha$ cluster model picture has been successful not only in understanding the structure in the bound energy region but also anomalous large angle scattering, and Airy structure in nuclear prerainbows and rainbows over a wide range of scattering energies in a unified way, as evidenced typically by  $\alpha$+$^{16}$O and $\alpha$+$^{40}$Ca systems \cite{Michel1998,Ohkubo1998,Ohkubo2016}.
  The $\alpha$ cluster picture based on crystallinity has been   confirmed from negative energy to high positive energy.

\par
  The  localized   three $\alpha$ cluster structure  of $^{12}$C, Fig.~1(a), has been  thoroughly  studied using    cluster models  based on   geometrical three $\alpha$ configurations,  including the      generator coordinate method (GCM)  with the Brink wave function \cite{Uegaki1977,Uegaki1979},
the resonating group method (RGM) \cite{Kamimura1977,Kamimura1981},
 semi-microscopic three $\alpha$ boson models  using
     the  orthogonality condition model (OCM) \cite{Kurokawa2007}
     and the Faddeev equation \cite{Fujiwara1976}.
 They   all  support  the  {\it  geometrical}  three $\alpha$ cluster structure in   $^{12}$C such as the   triangle geometry for the ground band,
  whose  precise   shape may be determined  by  experiment \cite{Fortunato2019}.

 \par
  On the other hand,  it  has recently been claimed that  typical $\alpha$ cluster structures  such as the ground band  of  $^{20}$Ne  and  $^{12}$C  can be understood  by  the completely opposing picture of  a  {\it nonlocalized cluster model} (NCM) without  any  geometrical configuration \cite{Zhou2012,Zhou2013,Zhou2014,Zhou2014PTP}. This  was originally proposed
 \cite{Tohsaki2001,Funaki2003}   to explain the  dilute gas-like $\alpha$ cluster structure of the Hoyle state in  $^{12}$C, which is
   considered to be a    BEC
    of  $\alpha$ clusters \cite{Matsumura2004,Itoh2004,Itoh2011,Freer2011,Nakamura2016,Nakamura2018,Katsuragi2018}.

\par
 The two  concepts  of  the {\it geometrical crystalline} cluster, Fig.~1(a),  and the  {\it nonlocalized}  cluster, Fig.~1(b),   are apparently incompatible with each other.  The challenge  is  to solve this puzzle  by
  reconciling the  two  exclusive  pictures because both pictures explain  the $\alpha$ cluster structure in the bound and quasi-bound energies almost equally well.

  \par
 The purpose of this paper is to show  that the crystalline cluster picture of Fig.~1(a) and the  nonlocalized cluster picture of Fig.~1(b) can be reconciled. We achieve this  by hypothesizing   that  the  $\alpha$ cluster structure has the   properties  shown in Fig.~1(d) of both crystallinity, Fig.~1(a), and   superfluidity, Fig.~1(c), simultaneously. We confirm this hypothesis for
  the   historically  most thoroughly studied $N$-$\alpha$ cluster nucleus,   $^{12}$C.
       It is essential  to rigorously treat the Nambu-Goldstone  (NG) zero mode  due to the spontaneous symmetry breaking
         of the global phase  in   finite systems  with a small number of particles. This  has not been respected in  traditional cluster models.

\begin{figure}[t]
\begin{center}
\includegraphics[width=8.6cm]{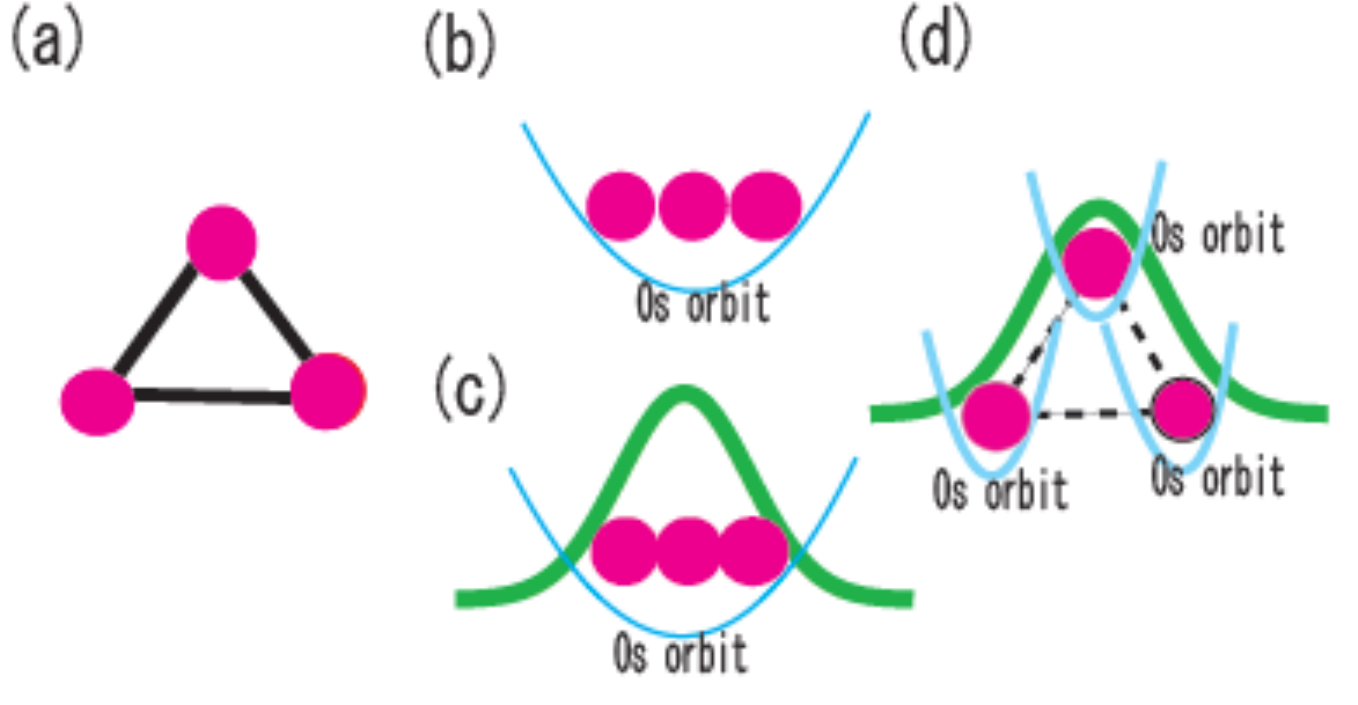}
\caption{ Illustrative   pictures of the  $\alpha$ cluster structure in $^{12}$C.  (a)   Geometrical  crystalline picture of the  three $\alpha$ clusters.
(b)   Nonlocalized cluster  picture of the  three  $\alpha$ clusters
in the  same 0s orbit of the  potential.
(c)   Superfluid  cluster model picture   of the  $\alpha$ clusters trapped in the  potential
  with the associated   coherent wave (broad  curve).
  (d)    Supersolid  picture   of the  crystalline $\alpha$ clusters   trapped in the  distinct (due to the Pauli principle) 0s-orbit of each potential associated with the  coherent  wave (broad curve).
}
\label{fig1}
\end{center}
\end{figure}

For this purpose, we use    a field theoretical superfluid cluster  model (SCM) in which the order parameter that satisfies the Gross-Pitaevskii   equation is defined and the number fluctuation of $\alpha$ clusters is taken into account.
 { The formulation was
    originally  developed
  to  study a gas-like BEC  state \cite{Nakamura2016,Nakamura2018,Katsuragi2018}.
 }
The model Hamiltonian for a bosonic field $\hpsi(x)$
$(x=(\bx,t))$ representing
the $\alpha$ cluster is given as follows:
\begin{align}
&\hat{H}=\intx \hpsi^\d(x) \left(-\frac{\nabla^2}{2m}+
V_\ex(\bx)- \mu \right) \hpsi(x) \notag\\
&\,\,+\frac12 \intxxd \hpsi^\d(x)
\hpsi^\d(x') U(|\bx-\bx'|) \hpsi(x') \hpsi(x) \,,
\label{Hamiltonian}
\end{align}
with   $V_\ex$   and $U (|\bm x -\bm x'|)$ being a mean field potential
in which  the $\alpha$ clusters are trapped
and   the residual  $\alpha$--$\alpha$ interaction, respectively.
We set $\hbar=c=1$. 

\par
When superfluidity  of $\alpha$ clusters occurs, i.e.
the global phase symmetry of $\hpsi$ is spontaneously broken, we decompose $\hpsi$
as $\hpsi(x)=\xi(r)+\hphi(x)$
where the c-number $\xi(r)=\bra{0} \hat\psi(x)
\ket{0}$ is an order parameter and is assumed to be real and isotropic.
To obtain the excitation spectrum, we need to solve
three coupled sets of  equations,  the Gross--Pitaevskii (GP) equation, the Bogoliubov-de Gennes (BdG) equations, and the zero-mode equation.
The GP equation  determines the order parameter, $\xi$, and
 is given by
\begin{equation}\label{eq:GP}
\left\{ -\frac{\nabla^2}{2m}+V_\ex(r) -\mu + V_H(r)
\right\} \xi(r) = 0 \,,
\end{equation}
where
$
    V_H(r) = \intxd U(|\bx-\bx'|)\xi^2(r')\,.
$  $\xi$ is normalized
with the  superfluid particle number $N_0$ as $\intx |\xi(r)|^2 = N_0$.
The superfluid density is given by $\rho_s$=$|\xi(r)|^2$/$N_0$.
The BdG equations  describe the collective oscillations of the superfluid
and are given by
\begin{align}
  \intxd
  \left(\begin{array}{cc}
        \Lc & \Mc \\
        -\Mc^* & -\Lc^*
      \end{array}\right)
  \left(\begin{array}{c}
    u_{\bn} \\
    v_{\bn}
   \end{array}\right)
  = \omega_\bn
    \left(\begin{array}{c}
        u_{\bn} \\
        v_{\bn}
\end{array}  \right),
\end{align}
where
\begin{align}
  \Mc(\bx, \bx')
  & = U(|\bx-\bx'|) \xi(r) \xi(r'),\, \\
  \Lc(\bx, \bx')
  & = \delta(\bx-\bx')
     \left\{ -\frac{\nabla^2}{2m}+V_\ex(r) -\mu + V_H(r)\right\}
     + \Mc(\bx, \bx')\,.
\end{align}
The index $\bm{n}=(n,\, \ell,\, m)$ stands for the main, azimuthal, and
magnetic quantum numbers. The eigenvalue $\omega_\bn$ is the excitation energy
of the BdG mode. For isotropic $\xi$, the BdG eigenfunctions can be taken to have separable forms,
$u_{\bm{n}}(\bm{x}) = \mathcal{U}_{n\ell}(r) Y_{\ell m}(\theta, \phi), \,
 v_{\bm{n}}(\bm{x}) = \mathcal{V}_{n\ell}(r) Y_{\ell m}(\theta, \phi).
$
We necessarily have an eigenfunction belonging
to a zero eigenvalue,  $(\xi(r), -\xi(r))^t$, and its adjoint function
$(\eta(r),\eta(r))^t$ is obtained as $\eta(r)= {\partial \xi(r)}/{\partial N_0}$.
The field operator is expanded as $\hphi(x) = -i{\hat Q}(t)\xi(r)+{\hat P}\eta(r)
+\sum_{\bn} \left\{{\hat a}u_\bn(\bx)
+{\hat a}^\dagger v^\ast_\bn(\bx) \right\}$
with the commutation relations $[{\hat Q}\,,\,{\hat P}]=i$ and
$[{\hat a}_{\bn}\,,\,{\hat a}^\dagger_{\bn'}]=
\delta_{\bn \bn'}$\,. The operator ${\hat a}_\bn$ is an annihilation operator
of the BdG mode, and the pair of canonical operators ${\hat Q}$ and ${\hat P}$, which are
called the NG or zero-mode operators, originate from the SSB of the global phase.
The  nonlinear Hamiltonian for $\hat{Q}$ and $\hat{P}$, $\hat{H}_u^{QP}$, whose explicit forms are given  in  Ref.~\cite{Katsuragi2018},   gives a discrete spectrum in  the zero--mode equation
$\hat H_u^{QP} \ket{\Psi_\nu} = E_\nu \ket{\Psi_\nu} 
(\nu=0,1,\cdots)\,$, just as  a one-dimensional quantum mechanical Hamiltonian with a binding potential does.
The total unperturbed Hamiltonian
is ${\hat H}_u=\hat H_u^{QP}
+\sum_{\bn} \omega_\bn {\hat a}_\bn^\dagger{\hat a}_\bn$.
The states that we consider are $\ket{\Psi_\nu}\ket{0}_{\rm ex}$ with
energy $E_\nu$, called the zero-mode state, and
$\ket{\Psi_0}{\hat a}^\dagger_\bn
\ket{0}_{\rm ex}$ with energy $\omega_\bn$,  called the BdG state,
where ${\hat a}_\bn \ket{0}_{\rm ex}=0$.

As in Refs.~\cite{Nakamura2016,Nakamura2018,Katsuragi2018}, we
take
 $ V_\ex(r)= m \Omega^2 r^2/2\,,$  and  $U (|\bm x -\bm x'|)$
 $ = V_r e^{-\mu_r^2 |\bm x -\bm x'|^2} - V_a e^{-\mu_a^2 |\bm x -\bm x'|^2}\,$, 
 with  $V_r$ and $V_a$ being  the     strength  parameters of the short-range  repulsive potential due to the Pauli principle  \cite{Ali1966} and long-range attractive  potential, respectively.
 The chemical potential is fixed by the specification of the superfluid particle number $N_0$.
We identify  the ground state   as the vacuum $\ket{\Psi_0}\ket{0}_{\rm ex}$.  The
range parameters $\mu_a$  and $\mu_r$ are fixed to the values $0.475$ ${\rm fm}^{-1}$ and  $0.7$ ${\rm fm}^{-1}$ in Ref.   \cite{Ali1966}, respectively.
 The two potential parameters, $\Omega$, which controls the size of the system, and $ V_r$, which prevents  collapse of the condensate, are determined to be
 $\Omega=4.093$ MeV$/ \hbar$ and $ V_r$=610 MeV. These
 reproduce the experimental root mean square (rms) radius, 2.45 fm, of the ground state, $\ket{\Psi_0}\ket{0}_{\rm ex}$ and   the   energy  level of the $0_2^+$ Hoyle state, identified as the first excited   zero-mode  state $\ket{\Psi_1}\ket{0}_{\rm ex}$.

 \par
 Before proceeding to the calculated  results and discussions, we briefly mention what is newly developed  from  our  previous papers \cite{Nakamura2016,Nakamura2018,Katsuragi2018},  where we  focused on      gas-like dilute $\alpha$ cluster states  like the Hoyle state with a considerably large  condensation rate such as 70\%  and showed that they can be  understood  as a BEC of $\alpha$ clusters in the strict sense   that the global phase of the system is spontaneously broken. BEC occurs whatever the    condensation rate and the density distribution    if    the global phase is locked.  What is new in the present paper is that  SSB of the global phase of the system with  non-gas-like  $\alpha$ cluster structure  occurs stably  even under small condensation rate such as 5\%, which has never been  thought about before in  studies of  $\alpha$ cluster condensation in nuclei. 
 Furthermore, the negative parity states of $^{12}$C are also treated, whereas the previous studies  of gas-like BEC states \cite{Nakamura2016,Nakamura2018,Katsuragi2018} discussed only the positive parity states.
  We find that our theory  reproduces well  the observed  negative parity states  in  $^{12}$C,    $3^-$ at 9.64 MeV and $1^-$ 10.85 MeV,   for the first time from the viewpoint of superfluidity  of the $\alpha$ cluster structure. 

 \par
In Fig.~2   the energy levels calculated   using our SCM without any geometrical crystallinity, Fig.~1(c),  and assuming  a small     superfluid density (condensation rate)  of 5\%, i.e.  $N_0=0.05\times3$,  are  compared with the  experimental data \cite{ENSDF} and     other  $\alpha$ cluster  model calculations  based on the geometrical  crystalline picture, Fig.~1(a).
In the SCM   the  $J^\pi$=$2^+$, $4^+$, $3^-$, and  $1^-$ states emerge as  BdG
    mode excitations,  and the $0_2^+$ Hoyle state  appears as an NG zero-mode  excitation  on the superfluid  vacuum.
  The agreement of the  $3^-$ and  $1^-$ states with experiment is good.
 The agreement  of  the $2^+$  and    $4^+$  states  with experiment would  be improved  if   the deformation of
 $V_{\rm ex}$ were taken into account, since it would  shift the excitation energy of  the $2^+$  and    $4^+$  states  downward and  upward, respectively.
 The agreement of the SCM with  experiment is comparable to the   GCM and RGM calculations, both of which locate the rotational  band $2^+$  and  $4^+$  states with  an  equilateral
 triangular $\alpha$ cluster configuration \cite{Uegaki1977,Uegaki1979} considerably lower than the experimental data.
 In Fig.~2(c) the  $3^-$ and  $1^-$ states  have
the equilateral
and    non-equilateral  triangular configurations  of the  three $\alpha$  clusters, respectively \cite{Uegaki1977,Uegaki1979}.
 We consider
        that the $\alpha$ cluster structure involves   both properties of     geometrical  crystallinity and  superfluidity, which  evokes  Landau's  two-fluid model  (normal fluid and superfluid) of He II \cite{Brink2005} and the duality (particle and wave) of light, in which the  superfluidity and the  wave nature are both caused by the formation of  a  coherent         wave function (order parameter) due to the BEC of the bosons
        belonging to a  zero eigenvalue.

 \begin{figure}[t]
\begin{center}
\includegraphics[width=8.6cm]{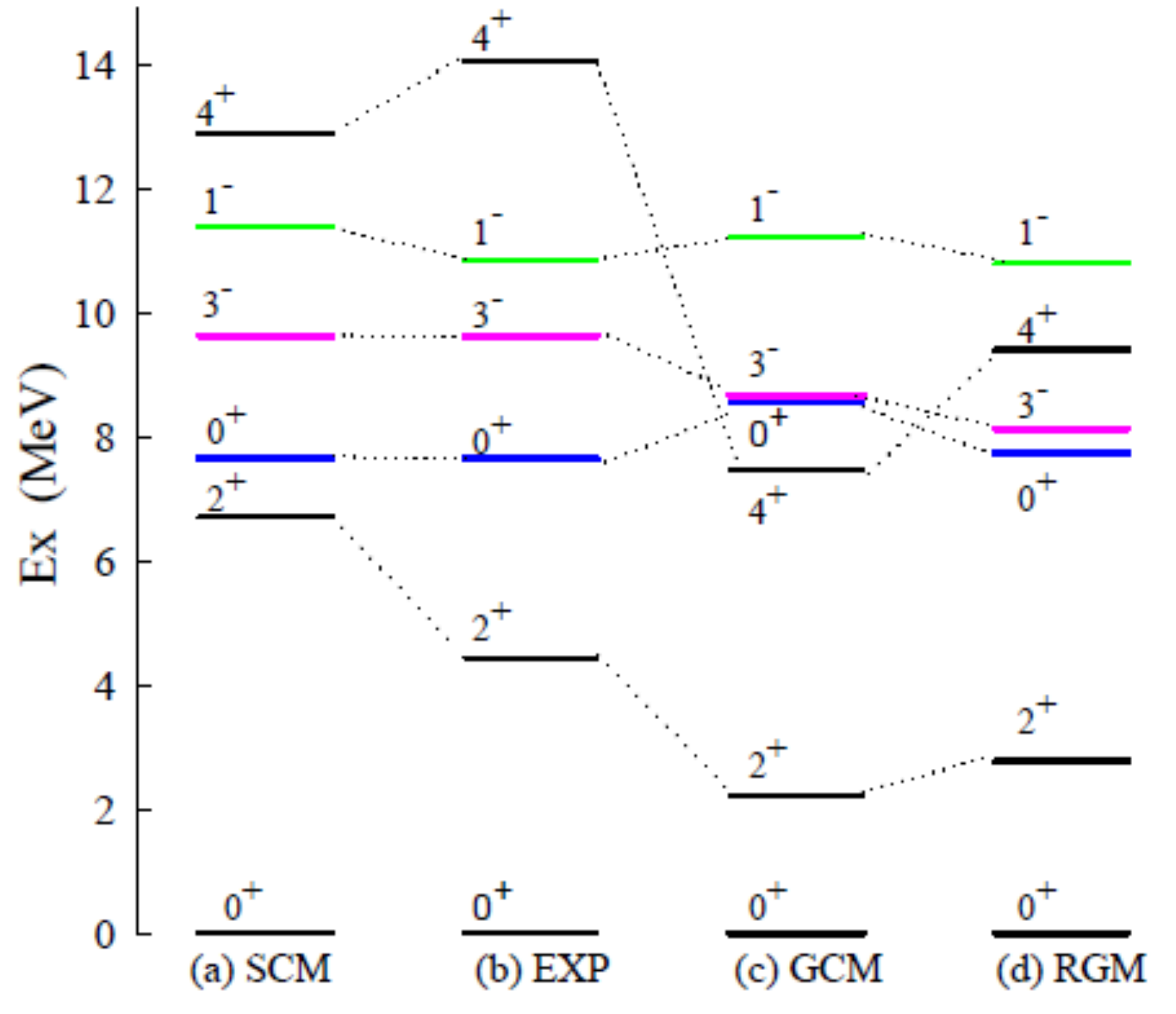}
\caption { Energy levels  for the
  $J^\pi$ states of $^{12}$C.  (a)  Superfluid   cluster model   (SCM) calculations. 
 (b) Experimental data  \cite{ENSDF}.  (c)   GCM \cite{Uegaki1977,Uegaki1979} and (d) RGM  \cite{Kamimura1977,Kamimura1981} three $\alpha$  calculations.
 }
\label{fig2}
\end{center}
\end{figure}

  \par
  In Fig.~3(a) the calculated eigenfunction
  $\xi(r)$   and its adjoint eigenfunction $\eta(r)$    are displayed.
   We see       that  the number fluctuation of the superfluid $\alpha$ clusters in the ground state, $\eta$,  is  highest near the surface region
   and decreases toward the inner and   outer regions.
 In Fig.~3(b)      $\rho_s$ and $\rho$ represent the probabilities of finding  the superfluid $\alpha$ clusters and nucleons, respectively.
    $\rho_s$ is  highest in the center of the nucleus and gradually decreases toward the surface region. The non-superfluid normal density may be defined as $\rho_n$$\equiv$ $\rho$-$\rho_s$.
    $\rho_s$ is  much smaller than      $\rho$. However, it is this small  superfluid density component    that causes  the  coherent     wave nature   of the  system.
   The  predisposition of the superfluid
     fraction  component $\rho_{s }$ in the ground state of $^{12}$C
    arises partly  due to  the orthogonality to      the BEC Hoyle state.

 \begin{figure}[t]
\begin{center}
\includegraphics[width=8.6cm]{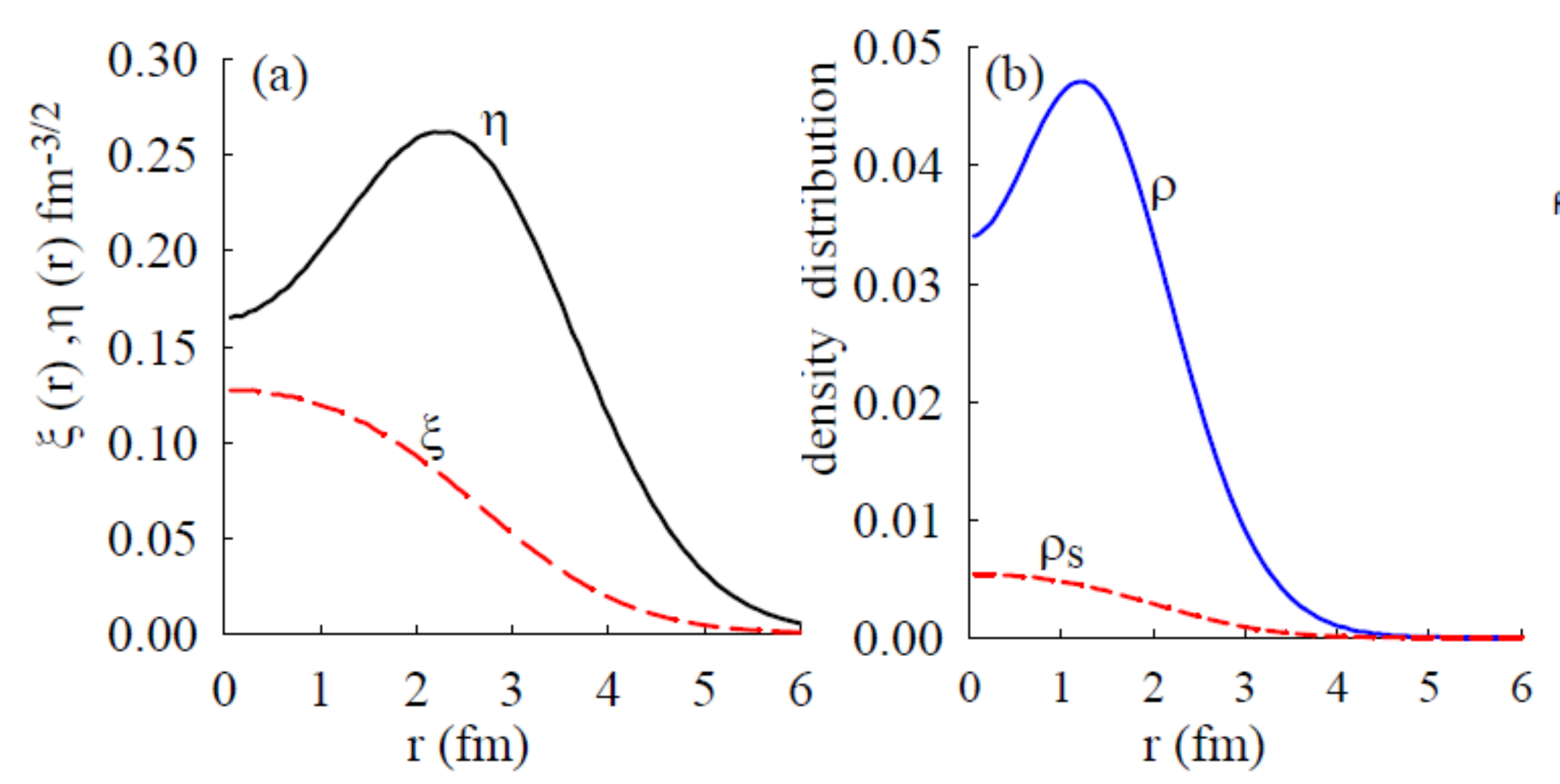}
\caption{(a) The calculated eigenfunction  (order parameter)  $\xi(r)$ (dashed line) and its adjoint eigenfunction $\eta(r)$ (solid line) for the ground state of $^{12}$C.  (b)   The  calculated  superfluid density distribution   $\rho_s$
  of the SCM  (dashed line) and the matter density distribution  $\rho$  of the RGM cluster model adapted from Refs. \cite{Kamimura1977,Kamimura1981} (solid line) for the ground state.
}
\label{fig:orderparameter}
\end{center}
\end{figure}
 
  \par
\begin{figure}[b]
\begin{center}
\includegraphics[width=7.6cm]{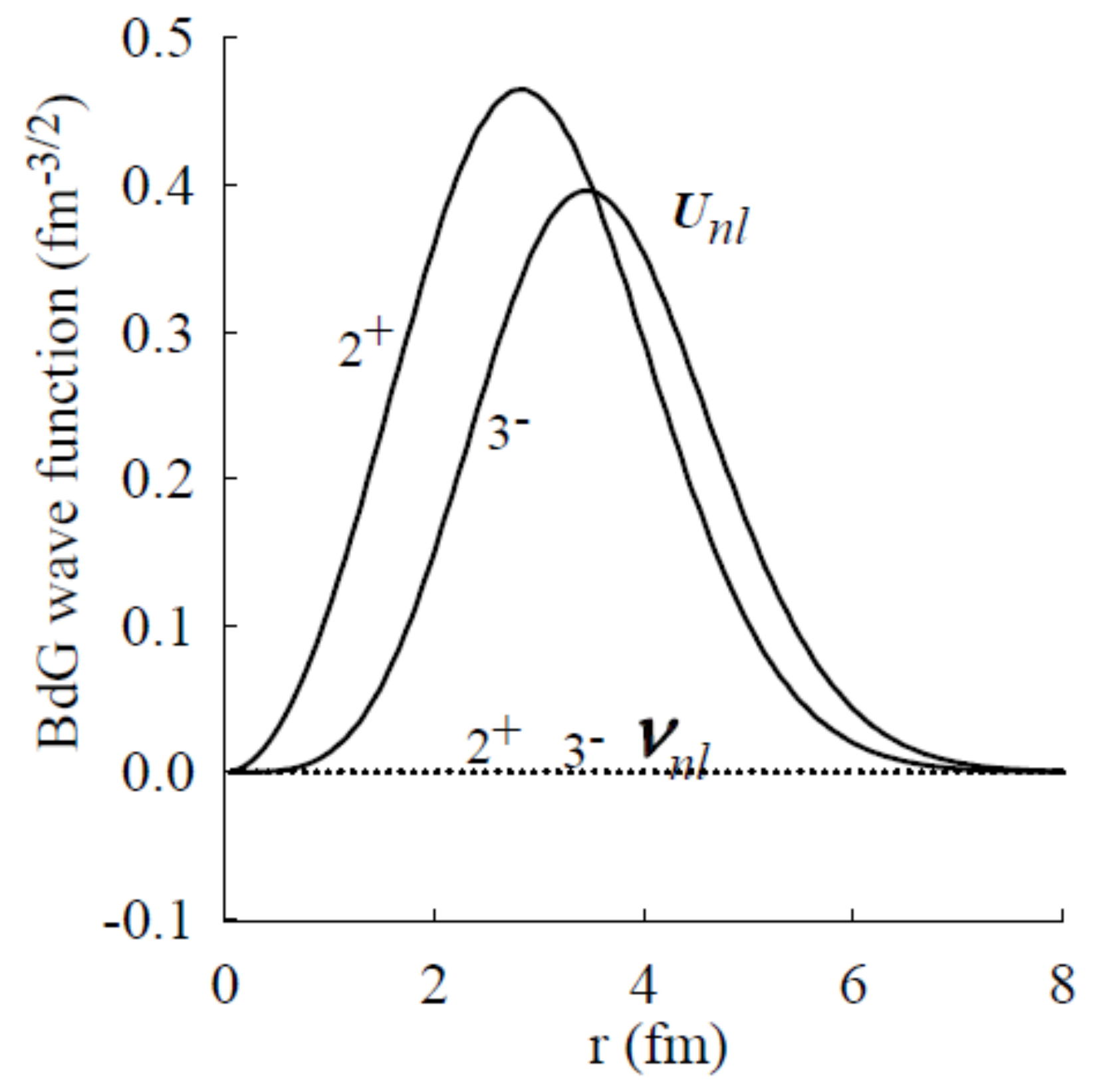}
\caption{Calculated  BdG wave functions $\mathcal{U}_{n\ell}(r)$ (solid lines) and $\mathcal{V}_{n\ell}(r)$  (dotted lines)   for the   $2^+$ ($n=0$, $\ell$= 2) and   $3^-$  ($n=0$, $\ell$= 3) states.
}
\label{fig:BdGwf}
\end{center}
\end{figure}

  In Fig.~\ref{fig:BdGwf}, the BdG wave functions $\mathcal{U}_{n\ell}(r)$ and $\mathcal{V}_{n\ell}(r)$     for the  $2^+$ and $3^-$ states are displayed.
The peak of $\mathcal{U}_{n\ell}(r)$
for $\ell \neq 0$  is located in the surface region because of the repulsive force between the  $\alpha$ clusters  and moves outward with increasing $\ell$ due to the centrifugal force.
The magnitude of $\mathcal{V}_{n\ell}(r)$ is  negligible  for the  $2^+$ and $3^-$ states, implying no Bogoliubov mixing  in these states  due to the  small condensation rate.

 \par
  We proceed to understand why   the  apparently exclusive pictures of  SCM, Fig.~1(c),  and GCM,  Fig.~1(a),    give  similar results.
  The geometrical structure in Fig.~1(a)
   has not previously been   considered  to be related to superfluidity     of $\alpha$ clusters.
 Also,   no attention has been paid  to the treatment of the  global phase
     of the wave function with a geometrical configuration.
    However,   we note  that in Fig.~1(d) bosons, $\alpha$ clusters of the Brink model in the GCM, which are  sitting  in the $0s$ state of distinct (due to the Pauli principle) harmonic oscillator potentials  and  are arranged with the geometrical configuration of Fig.~1(a),  can form a coherent wave. This is suggestive of    the optical lattice \cite{Yamamoto2013,Orso2006,Morsch2006,Bloch2008} in which trapped  cold atom bosons   form a  coherent condensed (superfluid) wave function.
    In fact, the de Broglie wavelength of each   $0s$ state $\alpha$ cluster with  very low energy
    is far larger than the geometrical distance $d$ between the $\alpha$ clusters. This  means that the phases of the waves are locked to form a  coherent wave function, i.e.  superfluidity (condensation) of the system.   This logic
      is general and  independent of the geometrical configuration and  number of   $\alpha$ clusters involved, $N$. Therefore  in principle, whatever the geometrical configuration, triangle, linear chain $N$-$\alpha$ cluster  ($N$=2, 3, 4,   $\cdots$)  or tetrahedron ($N$=4), trigonal bipyramid ($N=$5),   etc.,
   the geometrical $\alpha$ cluster structures have the potential  to form a coherent  wave function (superfluidity). Whether the   state  is superfluid  depends on    $\rho_s$, which   encapsulates  the structure and degree of  clustering.
  
   The  present study finds that the superfluid  ground state is stable with a condensation rate that is  5\%,  giving  similar energy levels to the GCM, RGM,  and experiment, as  shown in Fig.~2. This  strongly supports  the view of a  geometrical $\alpha$ cluster structure for the ground state with superfluid density that is sufficient  to form a coherent wave.
  We note  that the emergence of the  coherent wave function due to condensation  in nature is  possible even  if the   condensation rate is not  large.
  In fact, it was shown  through  systematic calculations \cite{Katsuragi2018} that BEC of  $\alpha$ clusters like the Hoyle state  occurs stably even under  a small condensation rate such as 20\%.
   We also note that the superfluidity of heavy  nuclei
    occurs due to the Cooper pairs generated  by a small number of nucleons near the Fermi surface    \cite{Brink2005}  as well the BEC of He II with a condensation rate of approximately  10\% \cite{Sears1982}.
It is useful to decompose the density distribution $\rho^{\rm GCM}$  due to  the GCM wave function $\Psi^{\rm GCM}$ based on Fig.~1(a)
  as  $\rho^{\rm GCM} = \rho_{s }^{\rm GCM} + \rho_{n}^{\rm GCM}$
where   $\rho_{s} ^{\rm GCM}$ is the superfluid density due to the   coherent wave function (order parameter) and $\rho_{n} ^{\rm GCM}$   is the noncondensed component.

\begin{figure}[t]
\begin{center}
\includegraphics[width=8.6cm]{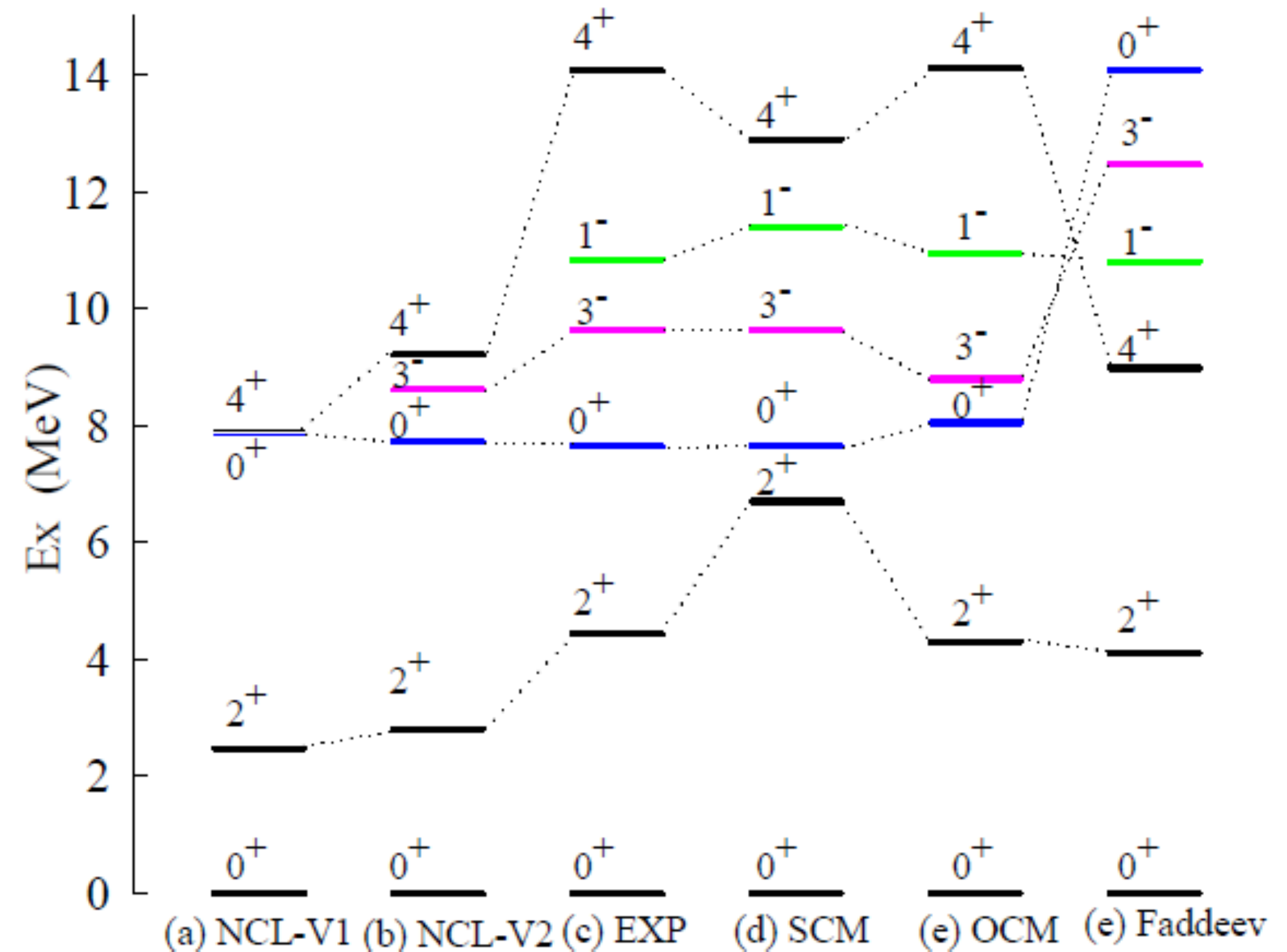}
\caption{  Energy levels of $^{12}$C  calculated in the nonlocalized cluster model  (NCM)   \cite{Funaki2005,Zhou2019} with  (a) Volkov  force No.1 and (b) Volkov  force No.2  are compared with (c) the  experimental energy levels
   \cite{ENSDF},  (d) the  superfluid cluster model (SCM)  calculations and  the  three boson model  calculations  using  (e) the   OCM \cite{Kurokawa2007} and   (f) the Faddeev model \cite{Fujiwara1976}.
}
\label{fig5}
\end{center}
\end{figure}

In Fig.~5 the energy levels of the NCM calculations \cite{Funaki2005,Zhou2019} with the Volkov force \cite{Volkov1965}   are compared with the SCM calculations, the experimental data and  the  three $\alpha$  boson model  calculations based on the geometrical crystalline picture  using  the   OCM \cite{Kurokawa2007} and   the Faddeev model \cite{Fujiwara1976}.
In Fig.~5(a) and (b)  $1^-$   is absent because it  was not  reported in Refs. \cite{Funaki2005,Zhou2019}.
 The NCM calculations reproduce the   experimental energy level ordering,  similar to the GCM, RGM, and SCM, although the $2^+$ and $4^+$ states are at  considerably lower energy than the experimental data.
Both the     NCM and  SCM      locate the $2^+$ state   deviated  from   experiment, downward and upward, respectively. The SCM  gives better  agreement   with the experimental   $4^+$ state. 
While  the ground band of the NCM  forms  a rotational  band, whose    moment of inertia    is much larger (1.8 times for V1 and 1.5 times for V2) than the experimental value,    the SCM  seems not to  form a rotational band.  This  is   understood as follows. While in the NCM calculations  the three $\alpha$ clusters are assumed to sit in   the   deformed  harmonic potential with the oscillator parameters  $B_x$=$B_y$$\neq$$B_z$   for $x$, $y$  and  $z$ directions   \cite{Funaki2005},  namely the trapping  (container) potential in Fig.~1(b) is deformed,     the trapping potential of the SCM  is assumed to be spherical in  the present calculations.
  If  the    SSB of  the rotational invariance is introduced in  the external potential $V_{\rm ex}$ in Eq.~(\ref{Hamiltonian}) of the SCM  by using  a deformed trapping harmonic oscillator potential   with $\Omega_x$=$\Omega_y$$\neq$$\Omega_z$ corresponding to $B_x$=$B_y$$\neq$$B_z$ in the NCM,  the rotational character of the  ground band would  be recovered. A reduction of  the  moment of inertia   due to the superfluid component  $\rho_s$ may be expected.

\par
The NCM  wave function    $\Psi^{\rm NCM}$  given by Eq.~(3) of Ref.\cite{Tohsaki2001} is obtained by constraining  the generating function
  of the GCM to the  Gaussian form  $f(\bm R)=\exp\left(-{\bm R^2}/{\beta^2}\right) $ \cite{Zhou2012}. It
 spans a   subspace of  the whole GCM Hilbert space. Physically,
   $\Psi^{\rm NCM}$           approximately extracts a
    condensate-like   component   from  the whole GCM wave function  $\Psi^{\rm GCM}$.
Although    $\Psi^{\rm NCM}$ with  all the $\alpha$  clusters  sitting in the $0s$ state as in Fig.~2(b) under  the antisymmetrization operator between the clusters  is a nonlocalized cluster wave function, it is not an exact  condensate wave function in the sense that the number fluctuations  of the $\alpha$ clusters
 are not   taken into account. It may be called a pseudo-condensed model
 because it involves the condensate component, which  can be dominant
 in the Hoyle state.
  The overlap of the GCM wave function   $\Psi^{\rm GCM}$ of Refs. \cite{Uegaki1977,Uegaki1979} with  $\Psi^{\rm NCM}$ is
about 0.93 for the ground state  \cite{Zhou2012}.
The ground state GCM (RGM) wave function, $\Psi^{\rm GCM}$($\Psi^{\rm RGM}$) of  Refs. \cite{Uegaki1977,Uegaki1979,Kamimura1977,Kamimura1981},  can be   represented well by a single $\Psi^{\rm NCM}({\bm \beta})$ with a large overlap of almost 100\% \cite{Zhou2014PTP}.
This means  that  the  GCM (RGM) wave functions based on the  geometrical picture,  Fig.~1(a),
  almost equivalently involve  the  nonlocalized cluster  of Fig.~1(b).  In other words,    $\Psi^{\rm GCM}$ has   duality  involving  both the   crystalline, Fig.~1(a), and  nonlocalized cluster, Fig.~1(b), nature simultaneously.
   Similar to $\rho_s^{\rm GCM}$,  $\rho^{\rm NCM}$  may be decomposed as $\rho^{\rm NCM}$=$\rho_{s }^{\rm NCM}$+ $\rho_{n }^{\rm NCM}$.
Physically $\rho_{s }^{\rm NCM}$ in Fig.~1(b)  corresponds to $\rho_{s }^{\rm SCM}$ in Fig.~1(c).
   It is now clear  that $\Psi^{\rm GCM}$ should  {\it not} be regarded simply as a crystalline  wave function as it appears  in  Brink's  wave function since  it  involves the  dual  nature of crystallinity and nonlocalized coherent  wave structure via $\rho_s$.  A crystalline $\alpha$ cluster  structure with superfluidity is called a supersolid.

Considering the duality of the $\alpha$ cluster structure, it is
  natural that    the energy spectrum of $^{12}$C  is
  described  by both the GCM and RGM based on the crystalline picture and  by the SCM and NCM based on the nonlocalized cluster picture with wave nature.
Since  the   coherent wave function is represented by
 $\Psi$ with a common phase $\Phi$,  i.e. $\Psi=|\Psi|\exp( i\Phi)$  \cite{Brink2005},  it is    natural that such a  wave function can be represented by a  single wave function
 if   the  parameter $\beta$, which determines the size parameter of the trapping  harmonic oscillator potential,  is properly chosen.  Because of the duality,  it is not surprising that not only in the 5-$\alpha$  nucleus $^{20}$Ne \cite{Zhou2012,Zhou2013} but also in many $\alpha$ cluster nuclei \cite{Zhou2014,Zhou2014PTP,Tohsaki2001,Funaki2003,%
 Funaki2009,Lyu2015,Lyu2016,%
 Funaki2018,Zhou2019,Zhou2020,Suhara2014}, which have been successfully   described by the crystalline picture, the wave functions are well represented by a single nonlocalized cluster wave function. It is a manifestation of the wave nature of the duality of the $\alpha$ cluster structure.

\par
 We can see the duality of crystallinity and nonlocalized cluster  in the  most typical  well-developed  crystalline dumbbell $\alpha$ cluster of $^8$Be.   The  solved GCM  wave function of the ground band, $\Psi^{\rm GCM}$,  is well represented with the overlaps, 0.96, 0.96, and 0.93  for the $0^+$, $2^+$, and $4^+$ states, respectively,  by a single Brink wave function  $\Psi^{\rm Brink}(R)$  with the
  distance parameter  $R$=3.5 fm \cite{Horiuchi1970}.  At the same time, $\Psi^{\rm GCM}$  is represented  by a single $\Psi^{\rm NCM}$ with an overlap of  almost 100\% \cite{Funaki2003}.   This means that the $\alpha$ cluster of  $^8$Be has  the duality of crystalline and  nonlocalized cluster nature.
$^8$Be, which is  a starting nucleus with a two-$\alpha$ linear chain structure in the Ikeda diagram \cite{Ikeda1968,Horiuchi1972} and in the extended Ikeda diagram \cite{Oertzen2001,Ohkubo1998},  can be considered a     prototype example of such  duality of the $\alpha$ cluster structure  in nuclei.

 \par
 To summarize,
  we have shown that  the  energy levels with the $\alpha$ cluster structure in $^{12}$C
  {as well as the rms radius of the ground state}, which have been understood to have a crystalline  structure described  well by     cluster models,  can  also be described  by a superfluid $\alpha$ cluster model.    The   $\alpha$ cluster structure is
 found
    to have    duality, simultaneously exhibiting properties that would intuitively  be considered
mutually exclusive: crystallinity and nonlocalized cluster structure.
 The  $\alpha$ cluster wave function based on a crystalline picture
    involves a  superfluid  component, whose coherent wave
represents the  nonlocalization of the wave function described  by  a nonlocalized cluster model.
Considering the $\alpha$ cluster structure as a supersolid    having  crystallinity and superfluidity, it is   not at all surprising   that it exhibits
  the    crystalline  and the nonlocalized pictures simultaneously.
{
The emergence of the low-lying collective Hoyle state  as  a Nambu-Goldstone mode is a manifestation of the SSB of   the global phase of the  ground state with crystallinity.
}
 The supersolidity of the $\alpha$ cluster structure  is considered to be   the first confirmed example of a stable  supersolid  in  nature.
\vspace{0.5cm}
\par 
{\bf Acknowledgments}

The authors thank  Dr. P. Suckling for careful reading of the manuscript and comments.
S.O.  thanks the Yukawa Institute for Theoretical Physics, Kyoto University for   the hospitality extended  during  a stay in  2019.
This work is supported in part by Grant-in-Aid for Scientific Research through grant no. 19K14619 provided by JSPS, and by Waseda University Grant for Special Research Projects (Project Number:2019Q-021).

\end{document}